\documentstyle[prl,aps,epsf]{revtex}

\newcommand{\bea}{\begin{eqnarray}}
\newcommand{\eea}{\end{eqnarray}}
\newcommand{\beq}{\begin{equation}}
\newcommand{\eeq}{\end{equation}}

\newcommand{\gsim}{{}^{\displaystyle >}_{\displaystyle\sim}}
\newcommand{\lsim}{{}^{\displaystyle <}_{\displaystyle\sim}}
\newcommand{\sun}{\odot}

\begin{document}

\draft

%%% Comment out following two lines if one column %%%
\twocolumn[\hsize\textwidth\columnwidth\hsize\csname
@twocolumnfalse\endcsname

%%%%%%%%%%%%%%%%%%%%%%%%%%%%%%%%%%%%%%%%%%%%%%%%%%%%%%%%%%%%%%%%%%
\title{Structure Formation in Non-singular Higher Curvature Cosmology}

\author{Shinsuke Kawai${}^{1,2}$
%\thanks{\tt kawai@phys.h.kyoto-u.ac.jp} 
  and Jiro Soda${}^{3}$
%\thanks{\tt jiro@phys.h.kyoto-u.ac.jp}
}
\address{
  ${}^{1}$
  Graduate School of Human and Environmental Studies, Kyoto University,
  Kyoto 606-8501, Japan, \\
  ${}^{2}$
  Linacre College and Department of Theoretical Physics, University of Oxford,
  Oxford OX1 3JA, UK, \\
  ${}^{3}$
  Department of Fundamental Sciences, FIHS, Kyoto University, 
  Kyoto 606-8501, Japan.}
\date{\today}
\maketitle

%%%%%%%%%%%%%%%%%%%%%%%%%%%%%%%%%%%%%%%%%%%%%%%%%%%%%%%%%%%%%%%%%%
%%%%% Abstract %%%%%
\begin{abstract}
We propose novel structure formation scenarios based on a non-singular 
higher curvature cosmological model. 
The model is motivated by the $R^2$ coupling of a scalar field appearing in
the string theory, and in our scenarios the universe has no beginning and no 
end.
We give two examples with explicit parameter values which are consistent 
with present observations of the cosmological structure. 
In the first example, the origin of structures are generated as the adiabatic 
perturbation during the chaotic inflation, while in the second the isocurvature
non-Gaussian perturbation from the superinflating era is responsible for the
structure.
In the second case it is possible to generate primordial supermassive
blackholes whose scale is comparable to what is expected in the galactic
nuclei.
\end{abstract}

%%%%%%%%%%%%%%%%%%%%%%%%%%%%%%%%%%%%%%%%%%%%%%%%%%%%%%%%%%%%%%%%%%%
\noindent
\pacs{PACS number(s): 98.80.Bp, 98.80.Cq, 04.20.Dw}
\keywords{singularity, superstring, cosmological perturbation}
%\preprint{
\vspace{-22pt}
\hspace{340pt}
KUCP137, gr-qc/9906046
%}

%%% Comment out if one column %%%
\vskip2pc]
%%%%%%%%%%%%%%%%%%%%%%%%%%%%%%%%%%%%%%%%%%%%%%%%%%%%%%%%%%%%%%%%%%%
%%%%% Body of the Paper %%%%%

\section{Introduction}

Inflation is unarguably one of the essential ingredients of the modern 
cosmology. It gives a natural explanation to the flatness and homogeneity of 
the present universe, and the quantum fluctuation generated during the 
exponential expansion era explains the origin of the large scale structures
which we observe today.
The period of inflation, however, cannot continue from indefinite 
past\cite{vilenkin}, so the problem of the initial singularity remains 
unsolved. 

One possible mechanism of resolving the initial singularity is the
quantum tunneling scenario\cite{creation}, which is well-motivated by 
the quantum mechanical picture of the early universe. There are, however, 
other possibilities of constructing non-singular cosmological solutions,
obtained by introducing higher curvature terms which are ubiquitous in
quantum gravity and string theory. Construction of non-singular cosmological
models by resorting to generalized gravity theories deserves serious 
speculations since quantum gravitational effects are expected to dominate
in the very early universe. 

In this paper we shall concentrate on the higher curvature resolution of the
initial singularity, and propose a non-singular cosmological scenario
consistent with the current observations.
If the initial singularity were to be removed, the curvature appearing in
any effective action must stay finite in the course of the cosmological
evolutions\cite{limited}. 
In our model the initial singularity is avoided by the dominance of the 
Gauss-Bonnet term coupled to a scalar field. A merit of this model is
the absence of redundant degrees of freedom which may appear by introducing
higher curvature terms. The exact form of the $R^2$ coupling has been 
calculated in the context of string theory with certain compactification 
to four-dimensions\cite{r2coupling}, and non-singular cosmological solutions
based on this gravity theory are known.
The basic picture we propose in this paper is that the avoidance of the 
initial singularity takes place before the ordinary scenario of chaotic 
inflation.
We estimate the expected perturbation spectra in this scenario and 
discuss its implication to the current observations. 
We show that there are three different quantum fluctuation sources in our 
scenario:
the isocurvature fluctuations with parameter-dependent spectral index,
the adiabatic fluctuation of blue spectrum with index $n_\psi=10/3$,
and the adiabatic perturbation of flat spectrum from the chaotic inflation.
We shall see that the singularity avoidance in this model may leave its trace
as a perturbation of order unity in the observationally interesting scale,
which may end up now as the black holes at the center of the galaxies.

The rest of this paper is organized as follows: 
We briefly review the non-singular cosmological model proposed by Antoniadis,
et.al.\cite{art94} in Sec. II.
In Sec. III we present the outline of our non-singular cosmological scenario.
The perturbation spectra in this model is discussed in Sec. IV,
and the structure formation scenarios with explicit parameter values are 
discussed in Sec. V. 
In Sec. VI we conclude with some comments.

%%%%%%%%%%%%%%%%%%%%%%%%%%%%%%%%%%%%%%%%%%%%%%%%%%%%%%%%%%%%%%%%%%%

\section{Resolution of initial singularity by higher curvature}

We start with an action of higher curvature gravity motivated by a string 
loop correction, 
\beq
{\cal S}=\int dx^4 \sqrt{-g}
\left\{\frac 12 R-\frac 12 (\nabla\varphi)^2
  -\frac 18\xi(\varphi)R_{GB}^2+{\cal L}_{matter}
\right\},
\label{eqn:effaction}
\eeq
where 
$R_{GB}^2=R_{\mu\nu\kappa\lambda}R^{\mu\nu\kappa\lambda}
-4R_{\mu\nu}R^{\mu\nu}+R^2$. 
We assume the function $\xi(\varphi)$ to be the one appearing in the string 
loop corrections\cite{r2coupling},
\beq
\xi(\varphi)=-\ln[2e^\varphi\eta^4 (i e^\varphi)],
\label{eqn:ssxi}
\eeq
which is an even function of $\varphi$, and $\eta$ is the Dedekind $\eta$ 
function. 
The field $\varphi$ is the modulus appearing along with the 
orbifold compactification. 
We neglect the contribution ${\cal L}_{matter}$ from other matter contents 
for a moment.
The cosmological solutions from this action is studied in 
\cite{art94}.
Variation of the action (\ref{eqn:effaction}) using the flat
Friedmann-Robertson-Walker metric
%\beq
$
ds^2=-dt^2+a^2\delta_{ij}dx^idx^j,
$
%\eeq
where $a$ is the scale factor, yields the equations of motion
\bea
&&\dot\varphi^2=6 H^2 (1-H\dot\xi),
\label{eqn:bgeom1}\\
&&(2\dot H+5H^2)(1-H\dot\xi)+H^2(1-\ddot\xi)=0,
\label{eqn:bgeom2}\\
&&\ddot\varphi+3H\dot\varphi+3(\dot H+H^2)H^2\xi_{,\varphi}=0,
\label{eqn:bgeom3}
\eea
where $H$ denotes the Hubble parameter.
It can be shown that the system is invariant under 
$\varphi\leftrightarrow -\varphi$, and that the 
signs of $H$ and $\dot\varphi$ are conserved.
In the following we assume for definiteness the 
initial $\varphi$ to be negative and $\dot\varphi>0$. 

\begin{figure}
% \vspace*{0cm}
% \hspace*{0cm}
\epsfxsize=80mm
\epsfysize=40mm
\epsffile{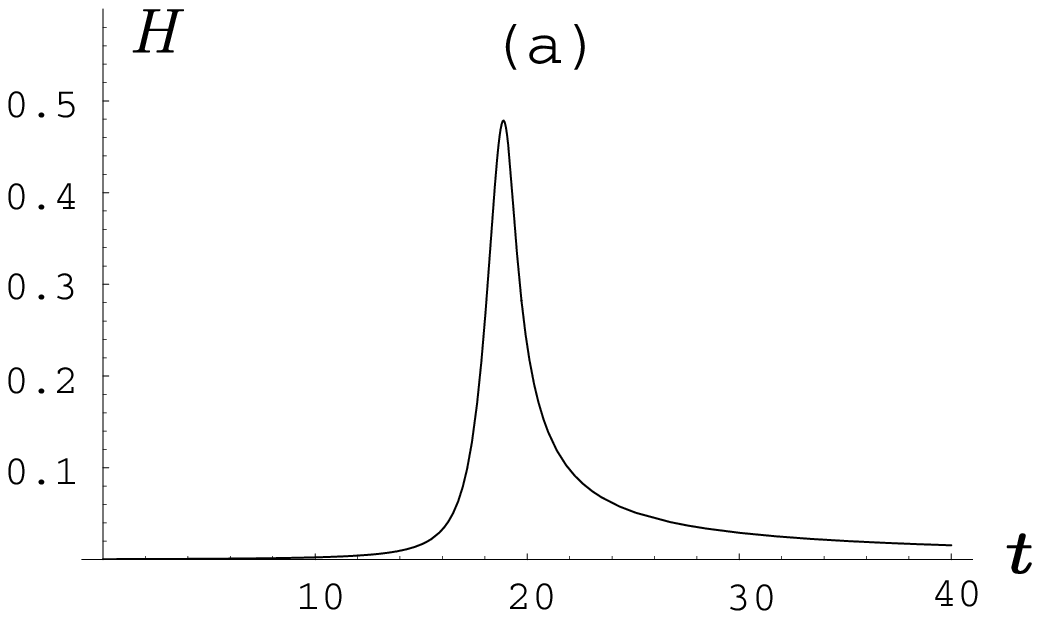}
%\vspace{-7mm}
\epsfxsize=80mm
\epsfysize=40mm
\epsffile{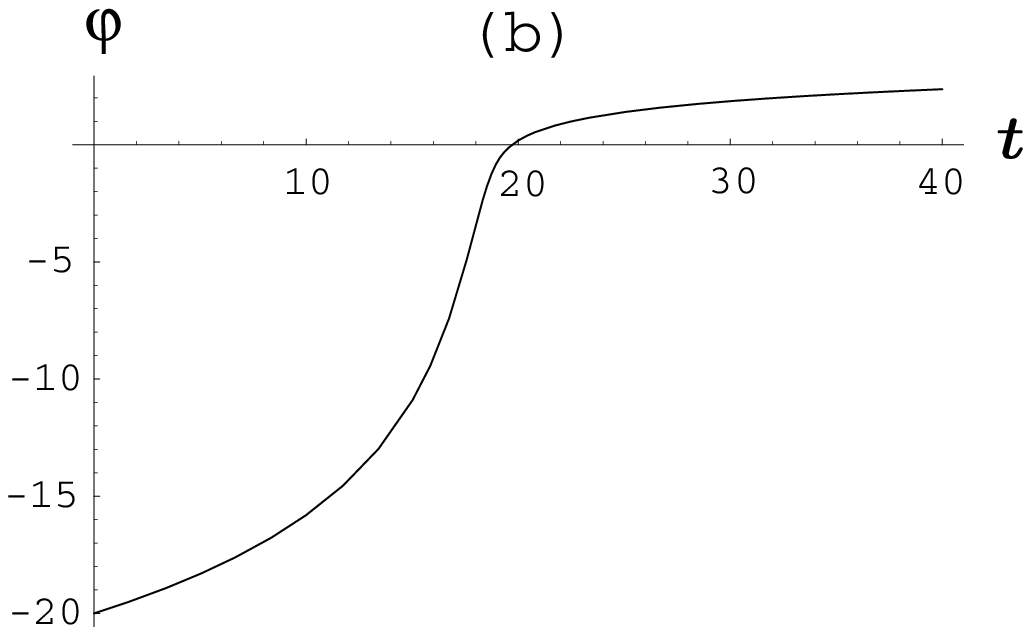}
%\vspace{-7mm}
\caption{The cosmological solution obtained from the action 
(\ref{eqn:effaction}). The universe makes a smooth transition from accelerating
phase to Friedmann-like phase. After the transition the scalar field
loses its kinetic energy and slows down.}
%\label{figure:phasediagram}
\end{figure}

Cosmologically interesting solution which leads to a Friedmann-like universe
in late time is such that $H>0$. This solution describes the
universe which starts from a Minkowski space with large enough $|\varphi|$ at
$t=-\infty$, gradually accelerates its expansion and increases the 
moduli kinetic energy as 
\beq
H\sim\frac{\omega_1}{|t|^2}, 
\hspace{1mm}\varphi\sim\varphi_p-5\ln|t|, 
\hspace{1mm}\dot\varphi\sim\frac{5}{|t|}, 
\hspace{1mm}a\sim a_p,
\label{eqn:paf}
\eeq
where $\omega_1$, $a_p$, and $\varphi_p$ are constants.
The constraint equation (\ref{eqn:bgeom1}) gives a relation 
$5e^{\varphi_p}=2\omega_1^3\pi$ between the 
initial values, so the initial degree of the freedom is only one,
apart from the overall scaling of $a_p$.
Note that, reflecting the special choice of the higher derivative terms
(the Gauss-Bonnet combination), there are no redundant degrees of freedom
which may make the phase space unphysically large.
Subsequently, the Hubble parameter reaches its maximal value $H=H_{\max}$ at 
$\varphi\sim0$, where moduli loses its kinetic energy and slows down.
After the Hubble peak, the universe naturally goes into a Friedmann-like 
phase, where
\beq
H\sim\frac{1}{3t}, 
\hspace{1mm}\varphi\sim\sqrt{\frac 23}\ln t, 
\hspace{1mm}a\sim a_f t^{\frac 13}.
\label{eqn:faf}
\eeq
Here, $a_f$ is a constant.
An example of the numerical solution is shown in Fig. 1.
In this model, there is no horizon problem since there is no particle horizon.
It is worth mentioning that this model is not plagued by a running scalar 
field, since the scalar field slows down in the Friedmann-like phase,
which is similar to the Damour-Polyakov mechanism\cite{damourpolyakov}. 

%%%%%%%%%%%%%%%%%%%%%%%%%%%%%%%%%%%%%%%%%%%%%%%%%%%%%%%%%%%%%%%%%%%

\section{Cosmological evolution}

Realistic cosmological models need to explain the thermal equilibrium in the 
early universe and the generation of present cosmological structures. 
Following the usual inflationary scenario, we assume that the reheating is 
caused by the coherent oscillation of the inflaton at the end of the chaotic
inflation which started shortly after the Hubble peak in the model described 
in the preceding section. This chaotic inflation is caused by
the large fluctuation around the Hubble peak which kicks up the inflaton
to sufficiently large expectation value.

As a component of the matter Lagrangian, we assume the inflaton Lagrangian 
of the form,
\beq
{\cal L}_{inflaton}=-\frac 12 (\nabla\chi)^2-V(\chi),
\eeq
where for simplicity we only consider the potential
\beq
V(\chi)=\frac 14 g\chi^4.
\eeq
After the Hubble parameter reaches its peak, the inflaton expectation value
becomes large due to the large fluctuation.
The universe region with sufficiently large $\chi$ satisfies
the slow-roll condition and inflates.
A natural initial condition for the chaotic inflation is estimated as
\beq
M_P^2 H_I^2 \simeq \frac 14 g\chi_I^4.
\label{eqn:initcond}
\eeq
The subscript $I$ denotes the value at the beginning of the 
chaotic inflation, which is supposed to be shortly after the Hubble peak. 
Then the e-folding number of the chaotic inflation is determined by the initial
value of the inflaton as
\beq
N\simeq\frac{\pi\chi_I^2}{M_P^2}.
\label{eqn:efolding}
\eeq
In our scenario there are four distinct eras. The first is the 
superinflationary era, which we call SI for short, with increasing Hubble 
parameter and accelerating modulus field. The second is the chaotic 
inflationary era, denoted CI, where the Hubble parameter stays constant and 
the scale factor increases exponentially. At the end of this chaotic inflation
the energy of the oscillating inflaton is converted into radiation, and 
the universe becomes radiation dominated (RD). As the radiation is redshifted
away more quickly than the matter, in the course of time the universe will be
matter dominated (MD).

%%%%%%%%%%%%%%%%%%%%%%%%%%%%%%%%%%%%%%%%%%%%%%%%%%%%%%%%%%%%%%%%%%%

\section{Perturbation spectra}

In our scenario, the inhomogeneity of the universe, which is expressed as
the perturbation on the homogeneous universe, is generated as the quantum
fluctuation inside the horizon. 
The quantum fluctuation is once red-shifted out of the horizon and 
classicalized in SI and CI eras, and reenters inside the horizon in RD and 
MD eras.
Since this inhomogeneity is supposed to be the seeds of the large scale
structure of the present universe, the spectrum of the fluctuation is crucial 
in predicting the large scale structure today and thus in discussing the 
plausibility of this scenario.
In this section, we speculate on the quantum fluctuation spectra generated 
during the SI and CI eras. 

\subsection{Adiabatic perturbation from superinflation}
In order to describe the fluctuations we expand the metric and the 
modulus up to first order of perturbations: 
\bea
ds^2&=&
-(1+2\phi)dt^2
+2aV_{|i}dt dx^i\nonumber\\
&&+a^2\left\{(1-2\psi)\delta_{ij}+2E_{|ij}\right\}dx^idx^j,\\
\label{eqn:metric}
\varphi&=&\varphi^{(0)}+\delta\varphi.
\eea
For the study of the fluctuations in SI era, it is convenient to use the 
uniform field gauge ($\delta\varphi=0$, $E=0$). 
In this gauge the perturbation of the Einstein equation reduces to one wave 
equation\cite{hwang}
\beq
\ddot\psi+\frac{\dot A}{A}\dot\psi+B\frac{\nabla^2}{a^2}\psi=0,
\label{eqn:spert}
\eeq
where
\bea
&&A=\frac{a^3\alpha(5\alpha^2-2\alpha+1)}{(3\alpha-1)^2},
\label{eqn:aterm}\\
&&B=\frac{(\alpha-1)^2(3\Gamma-2)}{5\alpha^2-2\alpha+1}-1.
\label{eqn:bterm}
\eea
Here, $\alpha$ and $\Gamma$ are defined as $\alpha=1-H\dot\xi$ and 
$\Gamma=-2\dot H/3 H^2$, respectively. 
As can be seen from (\ref{eqn:bgeom1}), $\alpha$ is (proportional to) the 
fraction of the modulus kinetic energy to the geometrical kinetic energy.
$\Gamma$ is the {\em effective adiabatic index} $\Gamma-1=-G^i{}_i/3G^0{}_0$,
where $G^\mu{}_\nu$ is the background Einstein tensor. They behave as 
$\alpha\sim t^2$, $\Gamma\sim t$ in the past asymptotic region
and $\alpha\sim 1$, $\Gamma\sim 2$ in the future asymptotic region (in the
sense of (\ref{eqn:faf})).
Since the scale factor is almost constant in the past region, 
$a^3\alpha$ rapidly decreases as the universe evolves towards the Hubble peak.
This leads to a decrease of the effective volume $A$ and the perturbation
behaves as if the universe is collapsing.
Note that the other variables $\phi$ and $V$ are 
written in term of $\psi$, as
\bea
&&\phi=\frac{2\alpha\dot\psi}{(1-3\alpha)H},\\
&&V=\frac{1}{a^2 \alpha}\int a\alpha\left[(3\Gamma-5)\psi-\phi\right]dt.
\eea

In the ordinary inflationary universe scenario, the fluctuation is 
quantum mechanically generated inside the horizon and is classicalized
as it is stretched to superhorizon.
We assume the similar picture in SI era as well as in CI era.
According to the conventional prescription, we regard $\psi$ as an 
Schr\"odinger operator and decompose it into mode functions
\beq
\hat\psi=\int d^3 k (2\pi)^{-3/2}
\left\{\hat a_k u_k(t)e^{ikx}+\hat a_k^\dag u_k^*(t)e^{-ikx}\right\},
\eeq
where $\hat a_k$ and $\hat a_k^\dag$ are annihilation and creation operators 
satisfying the commutation relation
$
\left[\hat a_k,\hat a_q^\dag\right]=\delta^3(k-q).
$
The mode functions $u_k$ and $u_k^*$ are determined so that $\hat\psi$
satisfies the equation (\ref{eqn:spert}). Using the past asymptotic 
solutions (\ref{eqn:paf}) in the superinflationary regime, 
the equation for $u_k$ becomes
\beq 
\ddot u_k+\frac 2t\dot u_k-\frac{4tk^2}{5\omega_1a_p^2}u_k=0.
\eeq
The canonically normalized solution of this equation which reduces to a 
positive frequency mode in the short wavelength limit ($k\gg a H$) will be
\beq
u_k=\frac{N_k}{t}
\left[
\mbox{Bi}
  \left(
    \left(\frac{4k^2}{5\omega_1a_p^2}
    \right)^{1/3}t
  \right)
+i\mbox{Ai}
  \left(
    \left(\frac{4k^2}{5\omega_1a_p^2}
    \right)^{1/3}t
  \right)\right],
\eeq
where Ai and Bi are the Airy functions and $N_k$ is the normalization factor
\beq
N_k=\pi^{1/2}
\left(\frac{k}{a_p}\right)^{-1/3}\left(\frac{3}{5a_p}
\right)^{3/2}
\left(\frac{5\omega_1^7}{4}
\right)^{1/6}.
\eeq
The long wavelength limit ($k\ll a H$) of this positive mode function 
represents the behavior of the perturbations outside the horizon. 
The power spectrum estimated shortly after the Hubble peak has the 
form
\beq
P_\psi=\frac{k^3}{2\pi^2}|\psi_k|^2\sim{\cal O}(10^{-1})
\times\omega_1^{\frac 53}
\left(\frac{k}{a_p}\right)^{\frac 73}\left(\frac{a_p}{a_*}\right)^6,
\label{eqn:psips}
\eeq
where the asterisk denotes the values estimated at (or more precisely, 
just after) the Hubble peak\footnote{
The adiabatic perturbation $\psi$ in the large scale limit does not stay 
constant in this model, but $\phi$ remains constant 
except in the vicinity of the Hubble peak.
We estimated the power spectrum (\ref{eqn:psips}) by using the constancy of 
$\phi$ on the background (\ref{eqn:paf}) and (\ref{eqn:faf}), and 
we discarded the constant mode of the long wavelength limit solution.
The classical evolution of perturbations is discussed in \cite{plb}.
}.
This is a steep blue spectrum with a spectral index 
$n_\psi=d\ln P_\psi / d\ln k+1=10/3$.
It should be noted that this expression includes no
independent variables, and is determined only by one background initial 
condition $\omega_1$. Thus the amplitude of the fluctuation is determined
unambiguously once the background is fixed. 

\subsection{Isocurvature perturbation from superinflation}
Apart from the adiabatic fluctuation described above, there is another
possible source of the fluctuation with significantly different spectrum in
SI era.
We consider a matter Lagrangian of an axionic field\cite{axion},
\beq
{\cal L}_{matter}=-\frac 12 e^{2\mu\varphi}(\nabla\sigma)^2.
\eeq
We decompose the axion field into its background part and a small 
perturbation, 
$\sigma=\sigma^{(0)}+\delta\sigma$,
and suppose $\sigma^{(0)}=0$ so that the background cosmological solution and
the perturbation $\psi$ are unaffected by the introduction of this axionic
field.
The equation for $\delta\sigma$ is
\beq
\delta\ddot\sigma+(3H+2\mu\dot\varphi)\delta\dot\sigma
-\frac{\nabla^2}{a^2}\delta\sigma=0,
\eeq
and the canonical quantization of $\delta\sigma$ using the past asymptotic
background solution (\ref{eqn:paf}) yields the spectrum\footnote{
Unlike $\psi$, $\delta\sigma$ stays constant outside the Hubble horizon as long
as $\mu>0$.}
\beq
P_{\delta\sigma}
=\frac{1}{8\pi^3}
\left(\frac{5}{2\omega_1^3\pi}\right)^{2\mu}
\left(\frac {k}{a_p}\right)^{2-10\mu}
2^{10\mu+1}
\Gamma(5\mu+\frac 12)^2.
\label{eqn:icps}
\eeq
The spectral index,
$
n_{\delta\sigma}=d\ln P_{\delta\sigma}/d\ln k+1=3-10\mu,
$
depends on the coupling $\mu$. 
For $\mu=0.2$, this gives a scale invariant flat spectrum.
It is noted that the statistics of this fluctuation is non-Gaussian.

\subsection{Perturbation from chaotic inflation}

Finally, there is a perturbation of flat $n=1$ spectrum arising from the
chaotic inflation. 
The amplitude of the density fluctuation at the horizon is estimated as
\beq
\zeta_\lambda\sim\left(\frac{H^2}{\dot\phi}\right)_\lambda
\simeq 2\sqrt\frac 23 g^{1/2}N_\lambda^{3/2}.
\label{eqn:zeta}
\eeq
Here, $N_\lambda$ is the e-folding number of the chaotic inflation since the 
scale $\lambda$ under consideration exits the horizon, until the time of 
reheating. 
Similarly, $\zeta_\lambda$ is the value for the corresponding comoving 
fluctuation scale $\lambda$. 

\subsection{Fluctuation spectrum in our scenario}
To sum up what have been calculated, we have three sources of perturbations
with different spectra:\\
1) Adiabatic perturbation of $n_{\psi}=10/3$ blue spectrum from
superinflationary era,\\
2) Isocurvature perturbation of $n_{\delta\sigma}=3-10\mu$ spectrum from 
superinflationary era,\\
3) Adiabatic perturbation of $n_\chi=1$ flat spectrum from chaotic inflationary
era.

\begin{figure}
% \vspace*{0cm}
% \hspace*{0cm}
\epsfxsize=80mm
\epsfysize=55mm
\epsffile{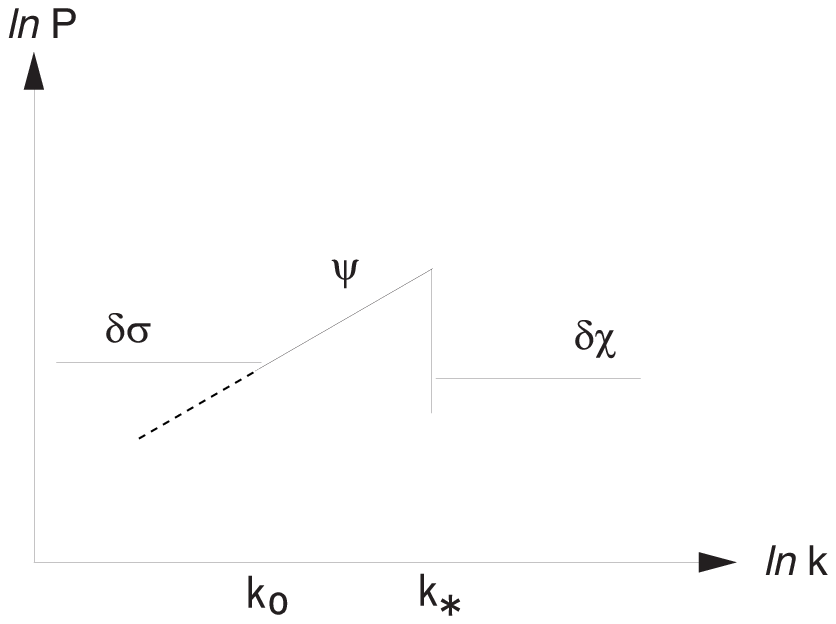}
%\vspace{-7mm}
\caption{Perturbation spectra in our scenario. $\delta\sigma$ is the
isocurvature perturbation and $\psi$ is the adiabatic perturbation generated 
in the superinflationary era. In this figure the perturbation $\delta\sigma$
is drawn as flat, but this can be tilted to red or blue, depending on
the axion type coupling constant $\mu$ which is chosen accordingly. 
The wavenumber corresponding to the horizon scale at the Hubble peak is denoted
by $k_*$. 
The other indicated wavenumber $k_o$ corresponds to the scale where the 
amplitude of $\psi$ becomes comparable to that of $\delta\sigma$.}
\label{figure:spectra}
\end{figure}

There might exist other sources of perturbations, such as the isocurvature mode
perturbation during CI era. However, we neglect them and assume these three
perturbations dominate for different scales and determine the whole primordial 
density perturbations. Since the universe is always expanding, the perturbation
generated in earlier time is expanded to larger scale at present.
Fig. \ref{figure:spectra} shows the expected perturbation spectra in our model,
where $\delta\sigma$ and $\psi$ is the isocurvature and adiabatic 
perturbations, respectively, in SI era and $\delta\chi$ is the adiabatic
perturbation in CI era. 
The inclination of $\delta\sigma$ can be tilted in either blue or
red direction depending on $\mu$. The figure shows the flat case
where $\mu=0.2$.

%%%%%%%%%%%%%%%%%%%%%%%%%%%%%%%%%%%%%%%%%%%%%%%%%%%%%%%%%%%%%%%%%%%

\section{Structure formation history}

Let us discuss the observational viability of our scenario.
In this model there are two distinct scales in fluctuations. One is the scale
corresponding to the horizon scale of the Hubble peak, which we denote by 
$\lambda_*$. 
The comoving wave number corresponding to this scale, $k_* =2\pi a/\lambda_*$, 
is shown in Fig. \ref{figure:spectra}. 
The other scale, $\lambda_o$, is where the amplitude of 
the blue adiabatic perturbation ($\psi$) is comparable to that of the 
isocurvature perturbation ($\delta\sigma$). 
The corresponding comoving wave number $k_o=2\pi a/\lambda_o$ is also shown 
in Fig. \ref{figure:spectra}. These scales are determined by the reheating 
temperature $T_{rh}$ and the initial inflaton value $\chi_I$, through the 
e-folding number during and after the chaotic inflation up to present.
Here we discuss two examples with explicit parameter values.

\begin{figure}
% \vspace*{0cm}
% \hspace*{0cm}
\epsfxsize=80mm
\epsfysize=55mm
\epsffile{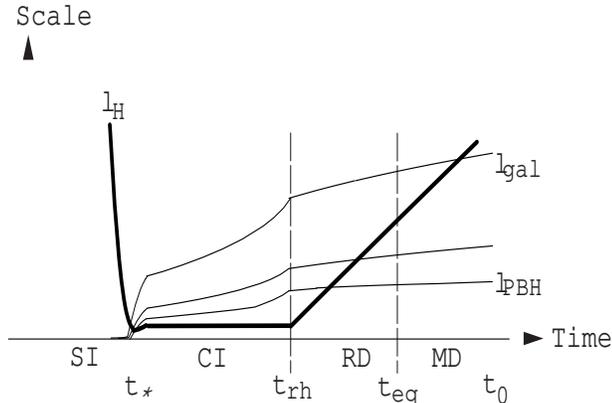}
%\vspace{-7mm}
\caption{A schematic picture of the Hubble horizon scale $\ell_H$ and the 
perturbation scales in our scenario.
Shortly after the Hubble peak the inflaton acquires a large expectation value
owing to the large fluctuations, and the chaotic inflation starts.
At the end of the chaotic inflation the universe is heated by the coherent
oscillation of the inflaton. See Sec. VB for the explanations of the scales
$\ell_{gal}$ and $\ell_{PBH}$.}
\label{figure:scales}
\end{figure}

\subsection{CI structure formation scenario}
We first consider the case where the scale $\lambda_*$ is large enough
compared to the horizon scale today, so that all of the observable structures
are generated during the CI era. 
In this case our scenario is nothing but that of the ordinary chaotic 
inflation, though the history before the onset of the chaotic inflation is 
included in our picture. 
Since the evidence of the singularity avoidance (the physics around the Hubble
peak) has been inflated away beyond the present horizon, this model is
consistent with our current observations only if the chaotic inflation
successfully takes place and generates appropriate amount of fluctuations.
In order to predict the fluctuation amplitude of 
${\cal O}(10^{-5})$, 
the coupling constant has to be 
$g\simeq{\cal O}(10^{-15})$.
If we assume the constant Hubble parameter during the CI to be
$10^{-7}M_P$, 
then the condition (\ref{eqn:initcond}) implies 
$\chi_I\simeq 4.9 M_P$.
From (\ref{eqn:efolding}), the e-folding number of the CI era will be
$N\simeq 74$ and we can see that the present horizon scale ($\sim \mbox{Gpc}$) 
is generated within the CI era.
Thus, with appropriate choice of parameters, the chaotic inflation naturally 
takes place and generates fluctuations with observationally acceptable 
amplitude.

\subsection{SI structure formation scenario}
As another possible scenario of structure formation, we can consider the 
case where both $\lambda_o$ and $\lambda_*$ are inside the horizon today.
In this case, what generates the observationally supported nearly
flat spectrum which evolves to the present large scale structure, 
is the isocurvature perturbation\cite{isocurvature} of SI era with appropriate 
choice of $\mu$
. 
Then the outstanding pointed spectrum of the $\psi$ perturbation and the 
flat spectrum of $\delta\chi$ perturbation, as is shown in 
Fig. \ref{figure:scales}, come inside the present horizon. This may fall into 
the scale of observational interest.
We choose, for example, the parameters as

\bea
&&H_*=0.1 M_P, \hspace{5pt}\mu=0.12, \hspace{5pt}g=10^{-12},
\hspace{5pt}\chi_I=4.6 M_P,
\nonumber\\
&& \hspace{5pt}T_{rh}=10^{12}~ \mbox{GeV}, 
\mbox{~and~} H_{CI}=H_I=10^{-5} M_P. 
\label{eqn:parameters}
\eea
Here, $H_{CI}$ is the energy scale of the chaotic inflation, which follows
from the condition (\ref{eqn:initcond}).
Using (\ref{eqn:efolding}), the e-folding number of the chaotic inflation is 
estimated as $N\sim 67$.  Taking account of the reheating temperature of
$10^{12}$ GeV and rather high energy scale $10^{-5}M_P$ during CI era, 
the large scale fluctuation of the universe which we observe 
today as CMB corresponds to the scale which exits the horizon during SI era.
From $\mu=0.12$, the isocurvature perturbation during the superinflationary
era has a blue tilted spectrum with $n_{\delta\sigma}=1.8$. 
The amplitude of the fluctuation for the present horizon scale is, from 
(\ref{eqn:icps}) with $\omega_1\simeq 0.1$ and $a_*/a_p\simeq 5$, estimated as 
$P_{\delta\sigma}^{1/2}\sim 10^{-3}$. 
 
The fluctuation generated during the chaotic inflation comes to scales 
smaller than $\mbox{kpc}$ today. If the amplitude of this small scale 
perturbation is too large, it will produce primordial blackholes. 
The primordial blackholes
of masses smaller than $10^{15}$ g will Hawking-evaporate within the age of the
universe, and if the radiation emitted through the evaporation
is too strong, it will spoil the observationally supported nucleosynthesis
scenario. Thus the amplitude of this small scale perturbation is constrained
by the amount of the primordial blackholes.
Assuming the spherical collapse model with Gaussian perturbation,
the primordial blackhole mass fraction at the time of their formation is 
estimated in terms of the perturbation amplitude $\delta (M)$ for the mass 
scale $M$ as\cite{carr}, 
\beq
\beta(M)\simeq\delta(M) \exp\left(-\frac{1}{18\delta^2(M)}\right).
\label{eqn:beta}
\eeq
The strongest constraint for the amount of the small blackholes is 
$\beta/(1-\beta)\lsim 10^{-26}$\cite{pbh}.
Using our choice of the parameters (\ref{eqn:parameters}), the amplitude of
the perturbation generated during CI era, from (\ref{eqn:zeta}), is
$\delta\sim 10^{-4}$. Substituting this into (\ref{eqn:beta}), we can see that
the small primordial blackholes are not overproduced.

The adiabatic perturbation of $n_{\psi}=10/3$, whose maximal amplitude is 
determined by (\ref{eqn:psips}), will produce certain amount of primordial
blackholes on its horizon reenter. The mass of these blackholes is determined,
according to the standard scenario of primordial blackhole formation, by the
Hubble horizon scale at the time of reenter. 
In our scenario the sharp peak of the fluctuation 
spectrum exits the horizon at $t_*$, inflated by the chaotic inflation and
reenters the horizon during the radiation dominated era of the universe.
For our parameters (\ref{eqn:parameters}), the scale of the perturbation peak
enters the horizon where $H\sim 10^{-2}\mbox{sec}{}^{-1}$. 
Then the mass of the typical primordial blackholes generated by the
$n_{\psi}=10/3$ spectrum is $M\sim 10^{7} M_\sun$.
The amount of these blackholes can be estimated by using (\ref{eqn:beta}).
We have $\delta\simeq P_{\psi}^{1/2}\simeq 0.05$ for our parameters, so
$\beta\simeq 10^{-10}$ at the time of horizon reenter. 
Translating this into the present fraction of the blackholes with respect to
the whole energy density of the universe, we have
$\Omega_{BH}\sim 10^{-6}$.
Since the spectrum of the adiabatic perturbation in SI era is steeply blue,
its contribution to the perturbation amplitude in larger scales is
negligible.

Recent observations suggest the existence of supermassive blackholes of
masses $10^7\sim10^9 M_\sun$ at the galactic center.
The amount of these blackhole is estimated using the radiation from the 
quasar, which is emitted along with the accretion of matter onto the
blackholes\cite{quasar}. 
The mass density of dead quasars is estimated as
$\rho_\bullet=u/\epsilon c^2\simeq 2.2\times10^5(0.1/\epsilon)M_\sun/
\mbox{Mpc}^3$,
which follows from the observed radiation density of quasars,
$u\sim 1.3\times10^{-15}\mbox{erg}/\mbox{cm}^3$.
In the expression of $\rho_\bullet$, $\epsilon$ is the efficiency of the 
radiation process.
Assuming that the mass density of dead quasars are comparable to that of 
blackholes, this indicates $\Omega_{BH}\sim 10^{-6}$.
Thus, our model with parameter values (\ref{eqn:parameters})
predicts both mass and density of the blackholes which are typically 
found in galactic centers.

\section{Discussions}
We have described a structure formation scenario based on a higher curvature 
non-singular cosmological model, and discussed two examples with explicit 
parameter values. Our rough estimates show that, with appropriate choice of 
parameters, this model can produce observationally acceptable amount of
perturbation amplitude in large scales, and even primordial blackholes in 
interesting mass scales.

%%% Hawking evaporation scenario
We have just assumed that the chaotic inflation, due to the large fluctuation, 
takes place after the Hubble peak,
and we have not discussed the physics of the onset of the chaotic inflation 
in detail. 
Of course there is a room to discuss other possibilities by including other
fields, etc., but this is beyond the scope of this paper.
Here, we comment on one possible alternative scenario
based on the study of the perturbation behavior.
Our background cosmological solution depends on the initial condition 
with only one degree of freedom.
In stead of $\omega_1$ in the discussion above, here we specify the background 
by $H_*$, the maximum Hubble parameter. 
For universes with small $H_*$, $B$ of (\ref{eqn:bterm})
remains negative throughout the evolution of the universe. 
For $H_*$ larger than unity in our unit, there appears shortly 
after the Hubble peak a period where $B$ becomes positive so the perturbation 
becomes unstable in short length\cite{plb}.
Since our unit is roughly Planckian, the Hubble parameter of the threshold of 
the instability is Planck scale. 
If this instability is physical, the perturbation inside the horizon will grow
exponentially in time and the region inside the horizon will collapse to form 
Planckian scale primordial black holes. 
Since small scale blackholes evaporate almost instantly, the universe will
be heated by the Hawking evaporation of these Planckian scale blackholes. 
Assuming that most of the energy density is converted to radiation during the 
unstable period, the reheating temperature of this scenario would be estimated 
as $ H^2\sim\frac 12 \dot\varphi^2\sim N_{eff}T_{rh}^4 $,
indicating the reheating temperature around the Planckian temperature.
Here, $N_{eff}$ is the number of particle species at the reheating epoch.
If this reheating takes place, we have to consider a thermal inflation rather
than the chaotic inflation in the following scenario. This inflation is
necessary in order to dilute unwanted relics coming out of the evaporating
blackholes. 
This scenario, however, might be somewhat speculative, and considering
that the action of our starting point (\ref{eqn:effaction}) is obtained by the 
loop correction in the string theory, a sensible interpretation would be that 
the effective theory breaks down for $H \gsim {\cal O}(1)$ and that this 
instability is unphysical.

\begin{figure}
% \vspace*{0cm}
% \hspace*{0cm}
\epsfxsize=80mm
\epsfysize=55mm
\epsffile{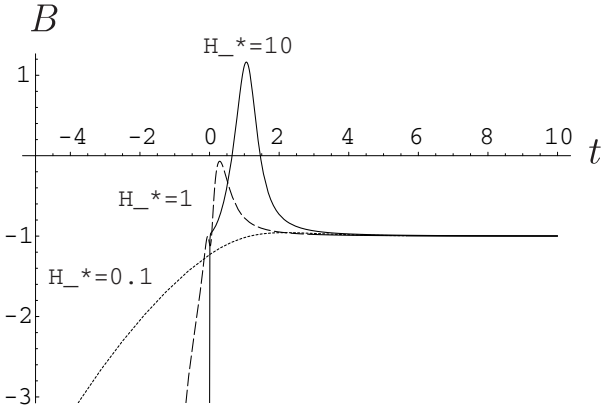}
%\vspace{-7mm}
\caption{
The time evolution of B (\ref{eqn:bterm}) for different background solutions 
which are specified by different values of $H_*$. 
Note that these are calculated on the numerical solutions of 
(\ref{eqn:bgeom1}-\ref{eqn:bgeom3}), and no further assumptions such as the 
continuation to the CI era are taken into account.
As can be seen from (\ref{eqn:spert}), positive B means the existence of the
instability for the scales smaller than the Hubble horizon.}
\label{figure:bterm}
\end{figure}

%%%%%%%%%%%%%%%%%%%%%%%%%%%%%%%%%%%%%%%%%%%%%%%%%%%%%%%%%%%%%%%%%%%

\acknowledgements

We appreciate the useful discussions with Atsushi Taruya.
Part of this work by J.S. is supported by the Grant-in-Aid for 
Scientific Research No. 10740118.

\appendix
\section{Formulae of cosmological perturbations}
The equations of motion derived from the effective action (\ref{eqn:effaction})
can be written as the following Einstein equation:
\beq
G^\mu{}_\nu=T_{\varphi}{}^\mu{}_\nu+T_{GB}{}^\mu{}_\nu+T_{matter}{}^\mu{}_\nu,
\eeq
where
\bea
G^\mu{}_\nu&=&R^\mu{}_\nu-\frac 12 R\delta^\mu{}_\nu,\\
T_{\varphi}{}^\mu{}_\nu &=& \varphi^{;\mu}\varphi_{;\nu}
 -\frac 12 \varphi^{;\alpha}\varphi_{;\alpha}
 \delta^\mu{}_\nu,\\
T_{GB}{}^\mu{}_\nu &=& 
(R^\mu{}_{\alpha\nu\beta}-R_{\alpha\beta}\delta^\mu{}_\nu
+R_{\alpha\nu}\delta^\mu{}_\beta)\xi^{;\alpha\beta}\nonumber\\
&&+G^\mu{}_\alpha\xi^{;\alpha}{}_{;\nu}-G^\mu{}_\nu\xi^{;\alpha}{}_{;\alpha},\\
T^{matter}_{\mu\nu} &=&\frac{1}{2\sqrt{-g}}
\frac{\delta{\cal L}_{matter}}{\delta g^{\mu\nu}}.
\eea
The non-vanishing components of $T_{GB}{}^\mu{}_\nu$ are
\bea
T_{GB}{}^0{}_0 &=& {}-3H^3\dot\xi,\\
T_{GB}{}^i{}_j &=& {}-\left[H^2\ddot\xi+2(\dot H+H^2)H\dot\xi\right]\delta^i{}_j.
\eea
Expanding $T_{GB}{}^\mu{}_\nu$ up to first order in perturbation gives
\bea
\delta T_{GB}{}^0{}_0 &=& \frac{\nabla^2}{a^2}
\left[H^2\delta\xi+3aH^2\dot\xi(V-a\dot E)-2H\dot\xi\psi\right]\nonumber\\
 &&+12H^3\dot\xi\phi+9H^2\dot\xi\dot\psi-3H^3\dot{\delta\xi},\\ 
\delta T_{GB}{}^0{}_i &=& \frac{\nabla_i}{a}
\left[H^2\dot{\delta\xi}-H^3\delta\xi
 -3H^2\dot\xi\phi-2H\dot\xi\dot\psi\right],\\
\delta T_{GB}{}^i{}_j &=& \frac{\nabla^i\nabla_j-\delta^i_j\nabla^2}{a^2}
\left[\ddot\xi\psi-H\dot\xi\phi-\frac 1a \left\{a^2 H\dot\xi(V-a\dot E)
\right\}\dot{}\right.\nonumber\\
&&-\left.(\dot H+H^2)\delta\xi\right]
+\left[2H\ddot\psi+2(\dot H+3H^2+H\ddot\xi)\dot\psi\right.\nonumber\\
&&+\left.3H^2\dot\phi+4H\left\{2\dot\xi(\dot H+H^2)+H\ddot\xi
\right\}\phi-H^2\ddot{\delta\xi}\right.\nonumber\\
&&-\left.2(H^3+H\dot H)\dot{\delta\xi}\right]\delta^i{}_j.
\eea
In the uniform field gauge, the equations of motion for the perturbations
(neglecting $T_{matter}^{\mu\nu}$) become:
\bea
&&3(3\alpha-1)H\dot\psi-6H^3\dot\xi\phi=2\alpha\frac{\nabla^2}{a^2}\psi
-(3\alpha-1)H\frac{\nabla^2}{a^2}V,\\
&&(3\alpha-1)H\phi+2\alpha\dot\psi=0,\\
&&\alpha\phi-(3\Gamma-5)\alpha\psi+\frac 1a (a^2\alpha V)\dot{}=0,\\
&&\ddot\psi+3H\dot\psi+H(1-\frac 32 H\dot\xi)\dot\phi\nonumber\\
&&\mbox{\hspace{20mm}}
=(H\dot\xi\dot\psi)\dot{}+(H\ddot\xi+2H^2\dot\xi+2\dot H\dot\xi)H\phi.
\eea
These four equations reduce to the wave equation (\ref{eqn:spert}).

%%%%%%%%%%%%%%%%%%%%%%%%%%%%%%%%%%%%%%%%%%%%%%%%%%%%%%%%%%%%%%%%%%%
 
%%%%%%%%%%%%%%%%%%%%%%%%%%%%%%%%%%%%%%%%%%%%%%%%%%%%%%%%%%%%%%%%%%%
\end{document}